%% file: salvaterra_nashville.tex
\documentclass[12pt]{article}
\usepackage{graphicx}

\newcommand\pubnumber{Article 7 in eConf C1304143}
\newcommand\pubdate{\today}

\def\iasf{INAF-IASF Milano, via Bassini 15, I-20133 Milano, ITALY}
\def\merate{INAF-OABr, via Bianchi 46, I-23807 Merate, ITALY}
\def\APC{APC Universit{\'e} Paris Diderot, 10 rue Domon et Dequet,
  F-75205 Paris, FRANCE}
\def\GEPI{GEPI – Observatoire de Paris, CNRS UMR 8111, Univ. Paris-Diderot, 5 Place Jules Jannsen, F-92190 Meudon, FRANCE}
\def\support{\footnote{ASI I/004/11/0 and the 2011 PRIN-INAF grant are acknowledged for financial support.}}

\def\Title#1{\begin{center} {\Large #1 } \end{center}}
\def\Author#1{\begin{center}{ \sc #1} \end{center}}
\def\Address#1{\begin{center}{ \it #1} \end{center}}

\newcommand\pubblock{\rightline{\begin{tabular}{l} \pubnumber\\
         \pubdate  \end{tabular}}}
\newenvironment{Abstract}{\begin{quotation}  }{\end{quotation}}
\newenvironment{Presented}{\begin{quotation} \begin{center} 
             PRESENTED AT\end{center}\bigskip 
      \begin{center}\begin{large}}{\end{large}\end{center} \end{quotation}}

\input econfmacros.tex

\begin{document}
\begin{titlepage}
\pubblock

\vfill
\Title{A Complete Sample of Long Bright {\it Swift} GRBs}
\vfill
\Author{R.~Salvaterra\support}
\Address{\iasf}
\Author{S.~Campana, S.~Covino, P.~D'Avanzo, G.~Ghirlanda,}
\Author{G.~Ghisellini, A.~Melandri, G.~Tagliaferri}
\Address{\merate}
\Author{L.~Nava}
\Address{\APC}
\Author{S.~Vergani}
\Address{\GEPI}
\vfill
\begin{Abstract}
Starting from the  {\it Swift} sample we define a complete sub-sample of 58 bright long Gamma Ray Bursts (GRB), 
55 of them ($95\%$) with a redshift determination, in order to characterize their properties. 
Our sample (BAT6) allows us to study the properties of the long GRB population
and their evolution with cosmic time. 
We focus in particular on the GRB luminosity function, on the spectral-energy correlations of their prompt emission,
on the nature of dark bursts, on possible correlations between the
prompt and the X-ray afterglow properties, and on the dust
extinction.
\end{Abstract}
\vfill
\begin{Presented}
Huntsville Gamma-Ray Burst Symposium\\
Nashville, USA,  April 15--19, 2013
\end{Presented}
\vfill
\end{titlepage}
\def\thefootnote{\fnsymbol{footnote}}
\setcounter{footnote}{0}

\section{Introduction}

Long Gamma Ray Bursts (GRBs) are powerful flashes of high energy
photons occurring at a rate of a few per day throughout the
universe. They are firmly associated with the death of massive stars
and therefore with star forming regions. 
Thanks to their brightness, GRBs can be detected up to extremely high
redshifts (so far we have a secure spectroscopic redshift of $z=8.2$
\cite{Salvaterra2009,Tanvir2009}) allowing us to study them and the
environment in which they explode
from the local universe to the epoch in which the first stars form.

\section{Sample selection}

\begin{figure}[thb]
\centering
\includegraphics[height=2.3in]{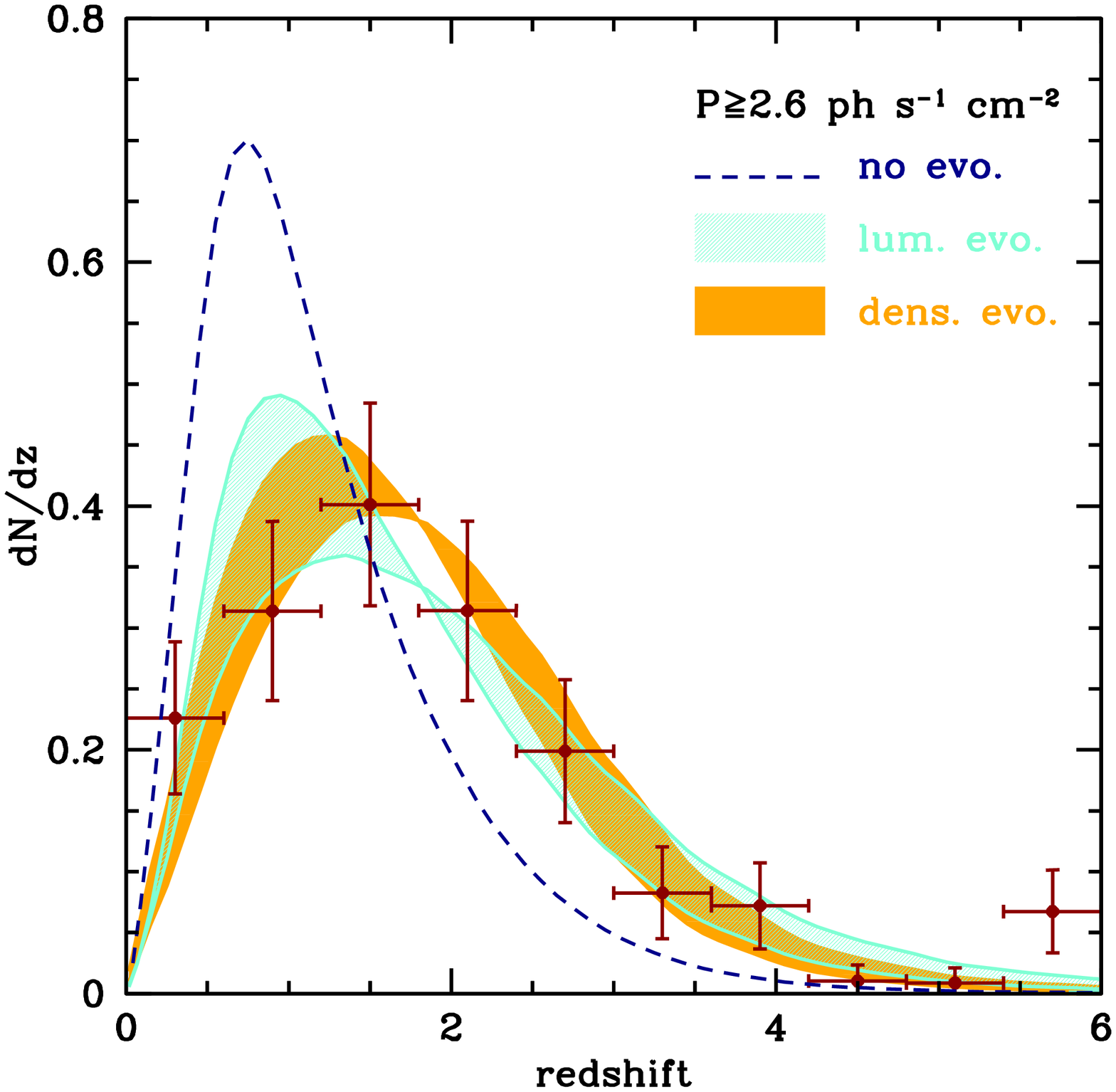}
\includegraphics[height=2.3in]{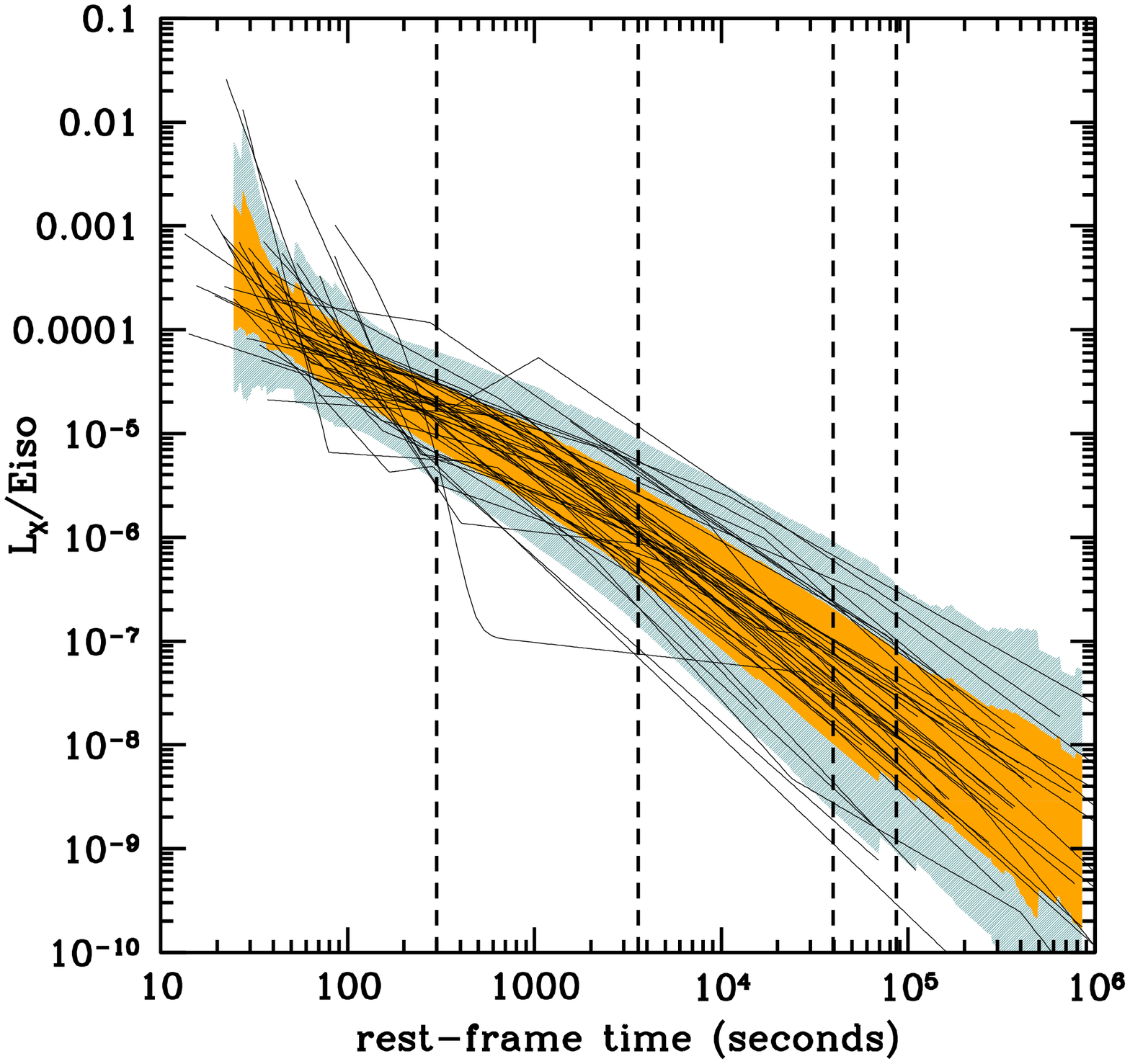}
\includegraphics[height=2.4in]{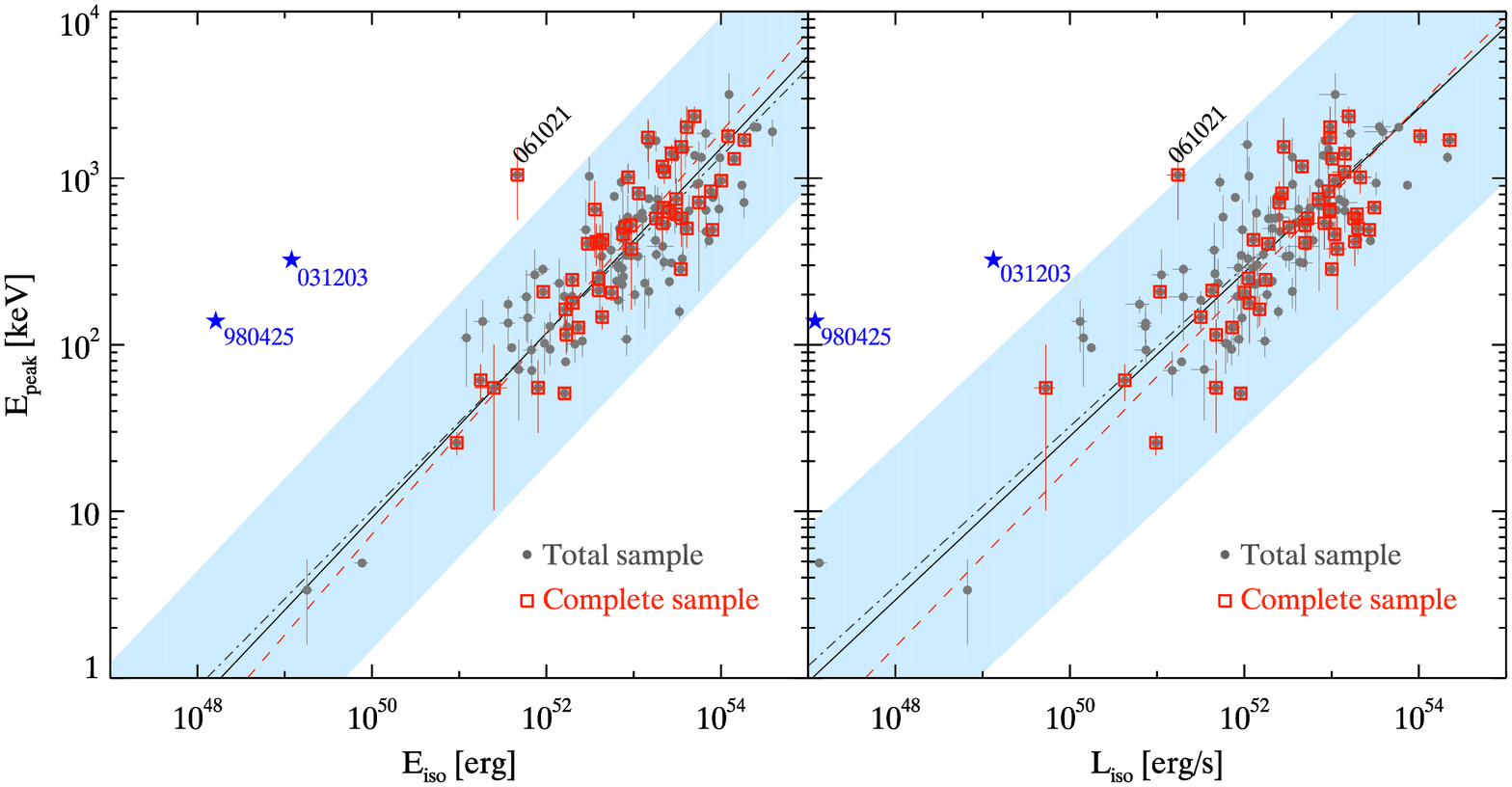}
\caption{Top left: redshift distribution of the BAT6
  sample \cite{Salvaterra2012}. Top-right: X-ray afterglow of the BAT6
  sample \cite{Davanzo2012}. Bottom:  {$E_{\rm peak}-E_{\rm iso}$} and
   {$E_{\rm peak}-L_{\rm iso}$} relation for the BAT6 sample \cite{Nava2012}}
\label{fig:1}
\end{figure}

About one third of the bursts detected by {\it Swift} has a measured redshift. While this represents an
enormous improvement with respect to the pre-{\it Swift} situation,
this is still too low to provide a complete sample in
redshift. 
Therefore we start from the criteria proposed by
\cite{Jakobsson2006} selecting those burst that have favorable
observing conditions for ground-based telescopes.
We then restrict to GRBs that are relatively bright in the 15-150 keV 
{\it Swift}/BAT band, i.e. with a 1-s peak photon flux $P\ge 2.6$ ph
s$^{-1}$ cm$^{-2}$ (BAT6 sample). This
corresponds to an instrument that is $\sim 6$ times less sensitive than
{\it Swift}, which give us a high level of confidence that all GRBs with a flux higher then our limit and that are inside the FOV of BAT
when they explode will be detected.  Therefore, our sample is complete with respect to our selection criteria 
and provide an unbiased view of the bright end of the GRB $\log N-\log
P$. 
58 GRBs match our selection criteria \cite{Salvaterra2012} and 55 of
them have a measured redshift (95\% completeness in redshift).
The mean (median) redshift is $1.84\pm 0.16$ ($1.60\pm 0.10$) with a long tail
extending at least up to $z=5.47$ (Fig.~\ref{fig:1}, left-panel).

\section{Results}

We briefly summarize here the results obtained on the basis of the
BAT6 sample.
\\

\noindent
{\bf GRB redshift distribution and luminosity
  function} \cite{Salvaterra2012}. We derive the GRB luminosity function (LF) by jointly fitting the observed differential number counts in 
the 50--300 keV band of BATSE and the observed redshift distribution of bursts in our sample.
As shown in Fig.~\ref{fig:1}, top-left panel, a no-evolution scenario
(dashed line, i.e. GRBs trace the cosmic SFR and their LF do not evolve) can not
reproduce the observed redshift distribution of the BAT6 sample. We
find that our data are well reproduced if we assume: (i) a luminosity evolution model, in which
the GRBs at higher redshift are typically brighter than those at lower
redshift with the typical luminosity increasing as $(1+z)^{2.3\pm 0.6}$; (ii) a density 
evolution model, in which the GRB formation rate increases with
redshift as $(1+z)^{1.7\pm 0.5}$ on the top of the known cosmic
evolution of the SFR, or (iii) GRBs form preferentially in low-metallicity
environments with $Z<0.3\;Z_\odot$.
Adopting these models we expect that $3-5$\% of bursts detected by
{\it Swift} should lie at $z>5$.

\noindent
{\bf Prompt emission correlations} \cite{Nava2012}. 
A strong correlation is found between the spectral peak energy, $E_{\rm
  peak}$, and the isotropic energy, $E_{\rm iso}$, or the isotropic
luminosity, $L_{\rm iso}$, for the bursts in the BAT6 sample
(see Fig.~\ref{fig:1}, bottom panel), with only 
one outlier, GRB061021, for the {$E_{\rm peak}-E_{\rm iso}$}. Their slopes, normalizations and dispersions are consistent with those
found with the whole sample of bursts with measured $z$ and {$E_{\rm
    peak}$}, and therefore
the biases present in the total sample commonly used to study these correlations do not affect
their properties.  We also find no evolution of the correlations with
$z$. Finally, we investigated the possible effects caused by the flux-limit selection in our complete sample on the {$E_{\rm peak}-L_{\rm iso}$}
correlation \cite{Ghirlanda2012}. If we assume that this correlation does not exist, we are unable to reproduce it as due to the flux
limit threshold of our complete sample. The null hypothesis can be rejected at more than 2.7 $\sigma$ level
of confidence.

\begin{figure}[t]
\centering
\includegraphics[height=2.3in]{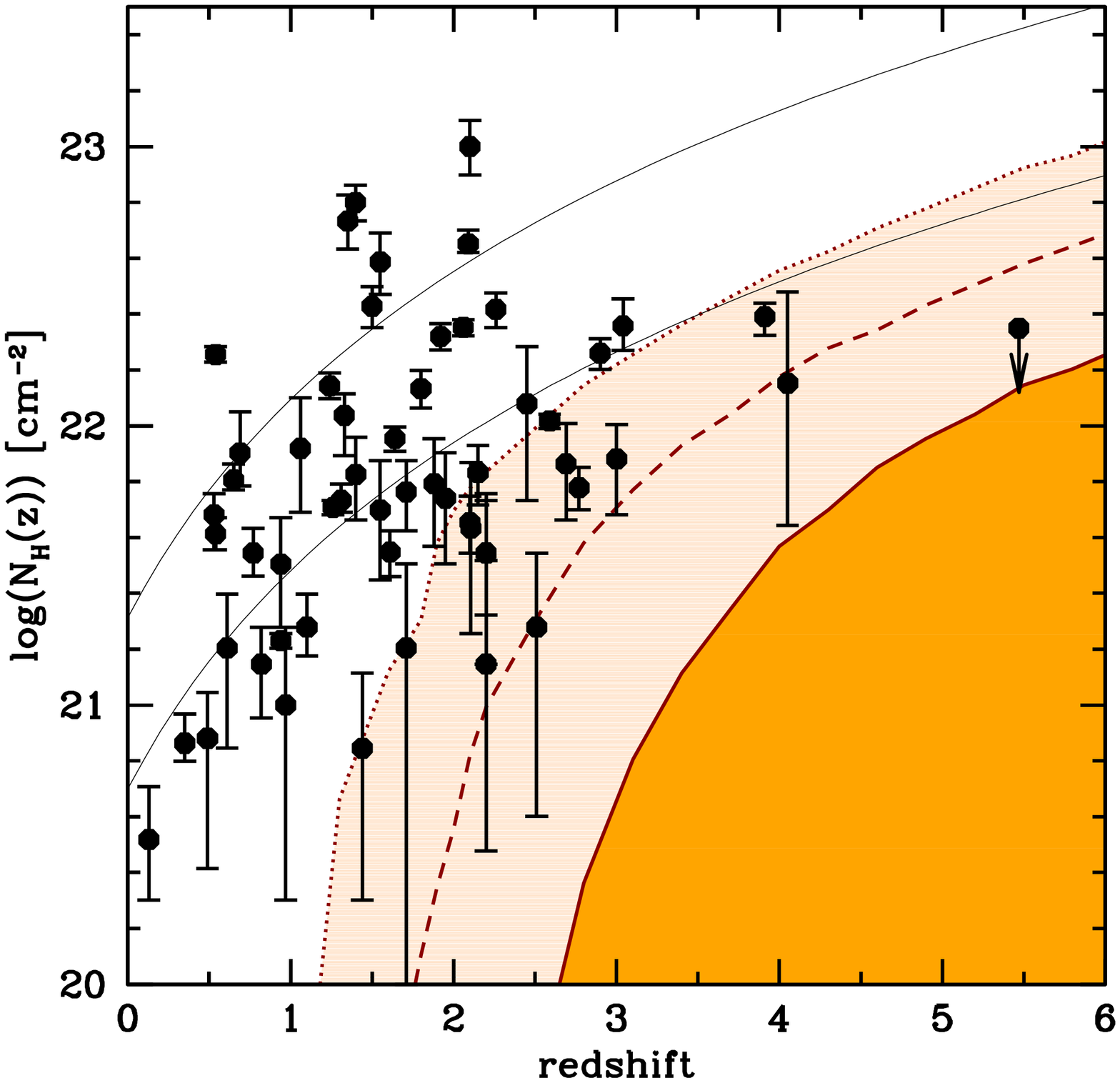}
\includegraphics[height=2.3in]{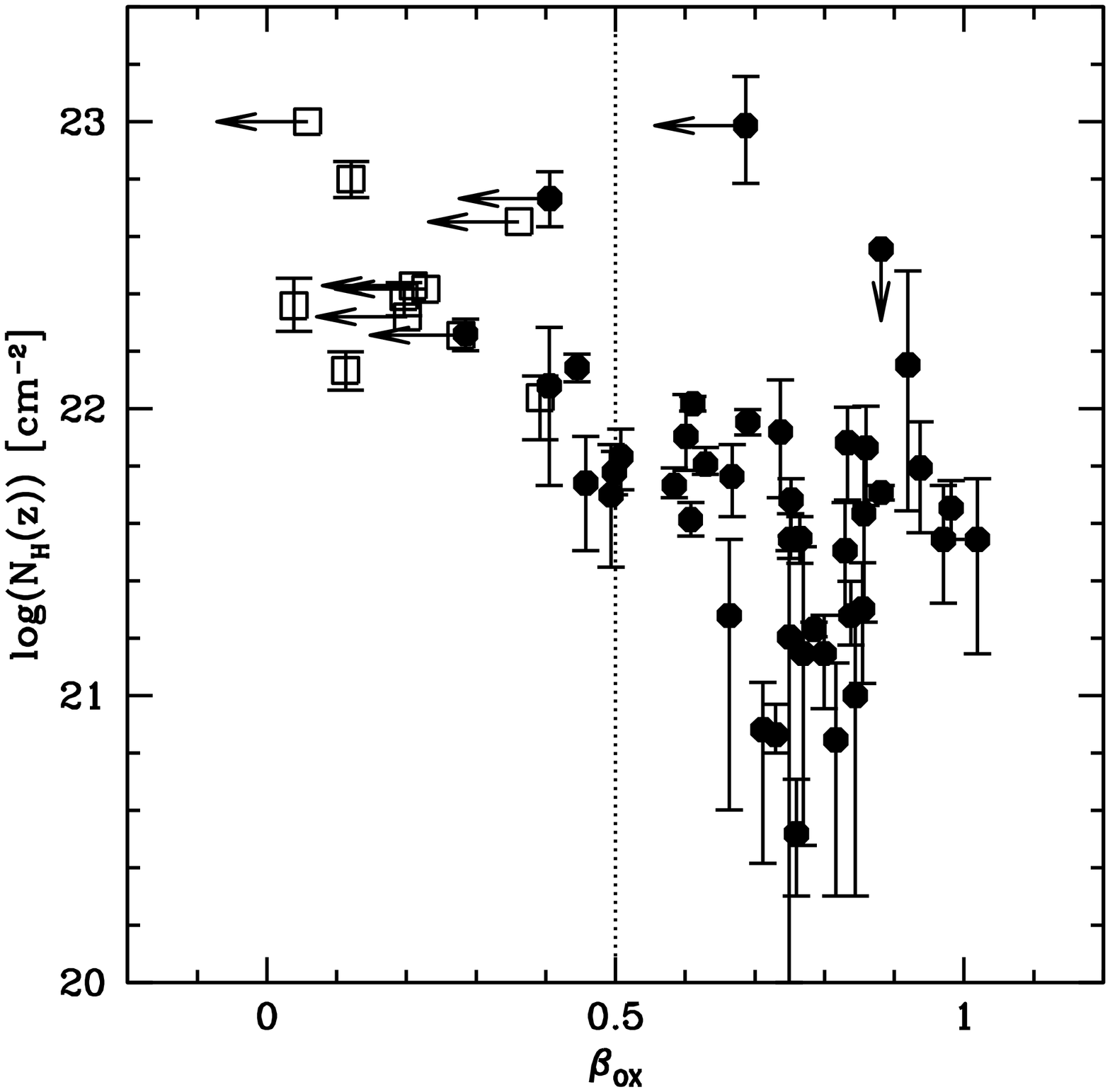}
\includegraphics[height=2.1in]{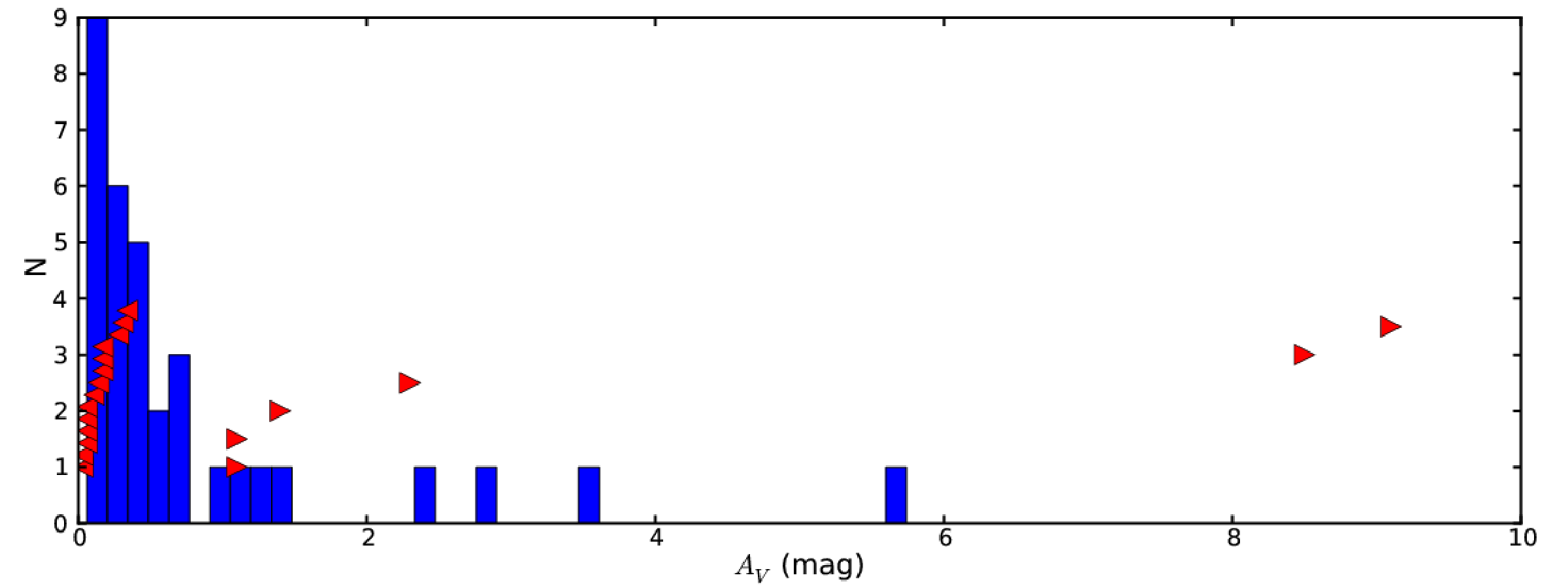}
\caption{Top left: X-ray intrinsic hydrogen column density as a
  function of redshift \cite{Campana2012}. Top right: relation between
  $\beta_{\rm ox}$ and
  $N_H$ \cite{Campana2012}. Bottom: $A_V$ distribution for the BAT6
  sample \cite{Covino2013}}
\label{fig:2}
\end{figure}

\noindent
{\bf X-ray afterglow} \cite{Davanzo2012}.
We find that the rest frame afterglow X--ray 
luminosity strongly correlates with $E_{\rm iso}$, $L_{\rm iso}$ and
$E_{\rm peak}$. These correlation decreases over time suggesting that the
X-ray light curve can be due 
to a combination of different components whose relative contribution and weight change with time, with the prompt and afterglow
emission dominating at early and late time, respectively. In particular, we found evidence that the plateau and the shallow decay
phases often observed in GRB X--ray light curves are powered by
activity from the central engine. The existence of the $L_{\rm
  X}-E_{\rm iso}$
correlation at late times (see Fig.~\ref{fig:1}, right panel) suggests a similar radiative efficiency among different bursts with on average
about 6\% of the total kinetic energy powering the prompt emission.

\noindent
{\bf X-ray absorption} \cite{Campana2012}.
For the full sample of X-ray afterglows, using the {\it Swift} X--ray Telescope data, we find that the distribution of
their intrinsic absorbing X--ray column densities has a mean value of $\log(N_H/{\rm cm^{-2}})=21.7\pm0.5$,
consistent with the one derived from the total sample of GRBs with
redshift \cite{Campana2010}. Contrarily with previous results, we find
that the low-$z$, high-$N_H$ region of the $N_H-z$ plane is now
populated by low redshift dark bursts. Still we find a lack of
high-$z$, low-$N_H$ GRBs suggesting a mild increase of the intrinsic
column density with redshift (Fig.~\ref{fig:2} left-panel). This can
be due to the contribution of intervening systems along the line of
sight (shadow area in Fig.~\ref{fig:2} left-panel; see
\cite{Campana2012} for details).

\noindent
{\bf Dark bursts} \cite{Melandri2012}.
``Dark"-GRBs are defined on the basis of the ratio between the optical
and the X-ray fluxes (or their upper limits). About one third of the
bursts in the BAT6 sample have optical-to-X-ray spectral index,
$\beta_{\rm ox}<0.5$ , i.e. they are dark according to the definition of \cite{Jakobsson2004}. Their redshift distribution and prompt
properties are very similar to those of the whole sample.
The major cause of the optically dark events is the
dust extinction  as also shown by the very tight correlation between
$\beta_{\rm ox}$ and $N_H$ \cite{Campana2012} (see Fig.~\ref{fig:2}
top-right panel). Indeed, this correlation shows that dark-GRBs form in a metal-rich
environment where dust must be present.

\noindent
{\bf Dust extinction} \cite{Covino2013}.
We find that 87\% of events in the BAT6 sample are absorbed by less than 2 mag (50\%
with $A_V<0.4$), the remaining being highly absorbed (Fig.~\ref{fig:2}
bottom panel). The latter appears to follow a different distribution
being inconsistent with a simple extrapolation of the low-extinction
events. No clear evolution of the dust extinction properties is
evident within the redshift range of our sample. As discussed before, dark bursts are
characterized on average by higher extinctions. Finally we find a
correlation between $A_V$ and $N_H$ although with a gas-to-dust ratio
well above that observed in Local Group environments.




\section{Conclusions}

We have presented here a complete sub-sample of bright {\it Swift}
long GRBs that is characterized by a high level of completeness in
redshift (95\%). This sample has been used to study the properties of
the long GRB population and their evolution with redshift. Apart from
the results highlighted here, the BAT6 sample has also been used
to study the distributions of the GRB jet opening
angle $\theta_{jet}$ and the bulk Lorentz factor $\Gamma_0$
\cite{Ghirlanda2012b}, the properties of precursors
\cite{Bernardini2013}, and the distribution of GRB afterglow radio
fluxes\cite{Ghirlanda2013}.


\end{document}

%% file: econfmacros.tex
\def\beq{\begin{equation}}
\def\eeq#1{\label{#1}\end{equation}}
\def\eeqn{\end{equation}}

\def\beqa{\begin{eqnarray}}
\def\eeqa#1{\label{#1}\end{eqnarray}}
\def\eeqan{\end{eqnarray}}

\let\bar=\overbar

\def\Dslash{\not{\hbox{\kern-4pt $D$}}}
\def\dslash{\not{\hbox{\kern-2pt $\del$}}}

\def\msb{{\bar{\ssstyle M \kern -1pt S}}}

